\documentclass[12pt,preprint]{aastex}

\shorttitle{The true nature of the alleged planetary nebula W16-185}
\shortauthors{Roman-Lopes et al.}

\begin{document}


\title{The true nature of the alleged planetary nebula W16-185\footnote{Based on observations %%@
made at Laborat\'orio Nacional de Astrof\'isica, Minist\'erio da Ci\^encia e Tecnologia, %%@
Brazil.}}


\author{A. Roman-Lopes\altaffilmark{1} and Z. Abraham\altaffilmark{1}}
%\affil{Instituto de Astronomia, Geof\'\i sica e Ci\^encias Atmosf\'ericas, Universidade de
%S\~ao Paulo \\ Rua do Mat\~ao 1226, 05508-900, S\~ao Paulo, SP, Brazil}

%\author{C. D. Biemesderfer\altaffilmark{4,5}}
%\affil{National Optical Astronomy Observatories, Tucson, AZ 85719}
%\email{aastex-help@aas.org}

%\and

%\author{R. J. Hanisch\altaffilmark{5}}
%\affil{Space Telescope Science Institute, Baltimore, MD 21218}


\altaffiltext{1}{Instituto de Astronomia, Geof\'\i sica e Ci\^encias Atmosf\'ericas, %%@
Universidade de
S\~ao Paulo \\ Rua do Mat\~ao 1226, 05508-900, S\~ao Paulo, SP, Brazil}
%\altaffiltext{2}{Society of Fellows, Harvard University.}
%\altaffiltext{3}{present address: Center for Astrophysics,
%    60 Garden Street, Cambridge, MA 02138}
%\altaffiltext{4}{Visiting Programmer, Space Telescope Science Institute}
%\altaffiltext{5}{Patron, Alonso's Bar and Grill}


\begin{abstract}
We report the discovery of a small cluster of massive stars embedded in a NIR nebula in the %%@
direction of the IRAS 15411-5352 point source, which is related to the alleged planetary nebula %%@
W16-185. The majority of the stars present large NIR excess characteristic of young stellar %%@
objects and have bright counterparts in the Spitzer IRAC images; the most luminous star (IRS1) %%@
is the NIR counterpart of the IRAS source. We found  very strong unresolved Br$\gamma$ emission %%@
at the IRS1 position and more diluted and extended emission across the continuum nebula. From %%@
the  sizes and  electron volume densities  we concluded that they represent ultracompact and %%@
compact HII regions, respectively. Comparing the Br$\gamma$ emission with the 7 mm free-free %%@
emission, we estimated that the visual extinction ranges between 14 and 20 mag. We found that %%@
only one star (IRS1) can provide the number of UV photons necessary to ionize the nebula.
\end{abstract}


\keywords{stars : formation -- stars: pre-main sequence -- infrared : stars -- ISM: HII regions %%@
-- ISM: dust, extinction}



\section{Introduction}

W16-185 was classified as a planetary nebula by \citet{wray66} but because of its proximity to %%@
the IRAS point source IRAS15411-5352, \citet{acker87} concluded that it is a compact HII region. 
Further investigation by \citet{nouma93}, through low-dispersion spectroscopy and narrow band %%@
CCD imaging in the H$\alpha$ line and the adjacent continuum, showed that the monochromatic %%@
H$\alpha$ image has its centroid displaced by $\Delta\alpha=11^{\prime\prime}$ and %%@
$\Delta\delta=-13^{\prime\prime}$ from the IRAS source  although it lies within its error %%@
ellipse.  
Even so, since  the object is located in the border between compact HII regions and PNe in the %%@
far infrared color-color diagram \citep{pott88}, they suggested that it was better described as %%@
a low-excitation planetary nebula. 

Although there is no point source in the 5 GHz PMN catalog \citep{wright94} in the direction of %%@
W16-185, the region can be considered part of the G326.7+0.6 radio source, associated with the %%@
RCW95 complex (Rodgers, Campbell \& Whiteoak 1960). \citet{goss70} mapped this region at 5GHz %%@
and found the maximum emission in the direction of the IRAS15408-5356 source and  an unresolved %%@
extension in the direction of W16-185. Strong CS(2-1) and NH$_3$(1,1) line emission were %%@
detected by Bronfman, Nyman \& May (1996) and  by \citet{roman02} for both IRAS sources. The %%@
radial velocities of the CS and NH$_3$ lines differ by less than 2 km s$^{-1}$ between the two %%@
regions, suggesting that they are at the same heliocentric distance.

\citet{roman04}  reported the existence of a rich cluster of OB stars centered at %%@
IRAS15408-5356. 
In this paper we report the detection of a  newly formed cluster of stars embedded in a small %%@
NIR nebula, associated to the IRAS15411-5352 source.
We observed the region in the $J$, $H$ and $K$ filters, as well as in the integrated Br$\gamma$ %%@
hydrogen line. 
The continuum subtracted Br$\gamma$ image shows two distinct regions, an unresolved ultra %%@
compact HII region (UCHII) and a more extended $10^{\prime\prime}$ compact HII region (CHII).
Using the Itapetinga radiotelescope at 43 GHz, with $2^\prime$.2 resolution, we were able to %%@
separate the radio source at the position of W16-185 from G326.7+0.6.

In Section 2 we describe the LNA and radio observations as well as the Spitzer/IRAC data. In %%@
Section 3 we identify the stellar cluster, the IRAS source and discuss the continuum and %%@
Br$\gamma$ emission from the nebula. In Section 4 we summarize our results. 

\section{Observations and data reduction}

\subsection{LNA Near Infrared data} \label{bozomath}

The imaging observations were performed in three observing runs: the  first two were performed %%@
in 2000 and 2003 May, using the Near Infrared Camera (CamIV) of the Laborat\'orio Nacional de %%@
Astrofisica (LNA), Brazil,  equipped with a HAWAII $1024\times 1024$ pixel HgCdTe array detector %%@
mounted on the 0.6 m Boller \& Chivens telescope. The images were made in the direction of IRAS %%@
point source 15411-5352, using the $\it{J}$ and $\it{H}$ broad band filters and the $\it{K}$ %%@
narrow band filters: $\it{C1}$,  centered in the continuum at $2.14 \mu$m  and Br$\gamma$, %%@
centered in the corresponding line.
Those observing missions were part of a survey \citep{abraham03} aimed to detection of the %%@
ionizing stars of  compact HII regions associated to IRAS point sources that are also strong CS %%@
emitters \citep{bronfman96}. 

The resulting data showed the presence of several embedded NIR sources as well as intense %%@
Br$\gamma$ emission in the direction of the IRAS source, however the $K$ band image resolution %%@
obtained with the 0.6 m Boller \& Chivens telescope (plate scale of 0.47 arcsec pixel$^{-1}$ and %%@
mean FWHM of about 2") limited our capability to study the stellar population in the direction %%@
of the nebula. In 2005 May we made new NIR imaging observations at LNA using the Near Infrared %%@
Camera attached to the 1.6 m Perkin-Elmer telescope.  The plate scale of this new observation %%@
was 0.24 arcsec pixel$^{-1}$ and the mean values of the  point-spread function FWHM were %%@
0.95$\arcsec$, 1.0$\arcsec$ and 1.1$\arcsec$ at the $\it{J}$, $\it{H}$ and $\it{K}$ ( $C1$ and %%@
$Br\gamma$ filters) images , respectively.
The total integration time for the $\it{J}$ and $\it{H}$ bands  were 330 s and  210 s %%@
respectively, and 1190 s for the $\it{C1}$ and $Br\gamma$ filters. The  completeness magnitude %%@
limits of 17.5 ($J$), 16.6 ($H$) and 15.0 ($C1$) were derived from the points at which the %%@
number $N(m)$ of detected sources  with magnitude $m$ deviate from a straight line in the log %%@
$(N)$ versus $m$ diagram, as can be seen in Figure 1.
The resulting frames were reduced using the IRAF reduction system and the Point Spread Function %%@
fitting photometry was performed using the DAOPHOT package \citep{stetson87}.
Details about the  calibration and reduction procedures can be found in \citet*{roman03}. %%@
Despite the strong nebular contribution present in the nebulae region, which could introduce a %%@
bias to the photometry, we are confident in our PSF fits because the star subtraction process %%@
gave good results, with no over-subtraction against the nebulosity.

To check the consistency of our photometry, we compared the LNA magnitudes with that of the %%@
2MASS survey. First we evaluated the color correction term from 43 isolated common stars outside %%@
of the infrared nebula, with good $H$ and $K$photometry in the 2MASS survey (flag A or B). As we %%@
can see from figure 2 the color term correction for the LNA photometric system is very small. In %%@
addition,  we made comparative magnitude diagrams for a sample of 101, 131 and 126 common stars %%@
in the $J$, $H$  and $K$ filters, respectively, which  are presented in Figure 3. 
We found a good linear relation between the two systems, with a slope of about 1 and a %%@
dispersion that increases with the magnitude. 

\subsection{Radio data} \label{bozomath}

The 7 mm radio observations were made with the 13.6 m radome-enclosed  Itapetinga %%@
radiotelescope, which has a beam size of about $2^\prime.2$ at this wavelength. The receiver is %%@
a $K$ band room-temperature mixer with 1-GHz double side-band, giving a system temperature  of %%@
about 700 K, and was operated in the total power mode.
The observations consisted in a series of 20 s right ascension scans  with $30^\prime$ %%@
amplitude. 
A linear baseline was subtracted from the data to eliminate the sky contribution. The  %%@
atmospheric opacity was calculated using the technique developed by \citet{abraham92} and the %%@
flux density was calibrated by observation of the standard  point source Virgo A. The quoted %%@
errors are the rms fluctuations of the flux densities, obtained by fitting a Gaussian function %%@
to the scans.

\subsection{Spitzer/IRAC data} \label{bozomath}

The Infrared Array Camera (IRAC) of the Spitzer Space %%@
Telescope\footnote{http://irsa.ipac.caltech.edu/Missions/spitzer.html} operates at four near to %%@
mid-infrared bands centered at 3.6, 4.5, 5.8 and 8.0 $\mu$m. The images used in this work are %%@
sections of the original frames of the SIRTF Galactic Plane Survey. The IRAC Post-Basic %%@
Calibrated Data (PBCD) of the region around IRAS15411-5352 were taken from the data archive of %%@
the Spitzer Science Center\footnote{http://ssc.spitzer.caltech.edu/}. Photometry of the Spitzer %%@
sources were performed using DAOPHOT package within IRAF, with an on source aperture radius of 3 %%@
pixels  and a sky annulus extending from 3 to 7 pixels. The aperture corrections and zero %%@
magnitudes fluxes were applied following the procedure described in the Spitzer/IRAC user %%@
manual.

\section{Results \& Discussion}

In Figure 4 (upper left side) we show the $\sim 3.9^\prime \times 3.5^\prime$  combined %%@
false-color image, made from the observed $\it{J}$ (green), $\it{H}$ (green) and $\it{C1}$ (red) %%@
images. We can see in the enlarged image (at the botton left side of that figure) several point %%@
sources embedded in a small and bright infrared nebula, suggesting the presence of a small %%@
cluster, which coincides with the position of the IRAS 15411-5352 error ellipse. From an %%@
inspection of the $R$ band image of the region (taken from the digitized all sky survey - %%@
DSS\footnote{http://archive.stsci.edu/cgi-bin/dss-form}) we verified that the near infrared %%@
nebula lies just bellow the $R$ image, implying that W16-185 is more extended than what is %%@
inferred from the optical data and that the extinction increases strongly to the south.

As mentioned before, W16-185 can be considered part of the RCW95 complex; in the  right part of  %%@
Figure 4 we show the 5 GHz contour map of the  radio source G326.6+0.6, associated to the RCW95 %%@
complex, obtained from  \citet{goss70}, in which the approximate position of the  observed NIR %%@
images is delimited.
Our 7 mm radio observations in the direction of IRAS15411-5352, with a $2.2^\prime$ beamwidth, %%@
were able to separate the source at this position; in a 30$^\prime$ scan at constant %%@
declination, it appeared as an isolated Gaussian profile centered at the IRAS source position %%@
with HPW of 3.4$^\prime$. The deconvolved angular size is 2.6$^\prime$  resulting in an %%@
integrated flux density of $5.2\pm 0.9$ Jy. The 7 mm complete map of the RCW95 region will be %%@
published elseware \citep{Barres05}.

\subsection{The stellar cluster} \label{bozomath}

In order to investigate quantitatively the star density in our images, we plotted in Figure 5 %%@
the positions of all sources detected above 3 $\sigma$ of the local sky in the LNA $K$ band %%@
image, labeled as IRAS15411-5352 panel. We can see a high concentration of stars near the IRAS %%@
coordinate (the "nebula" field); yet a high density of sources can result from localized low %%@
extinction as discussed by \citet{persi00}. On the other hand, young clusters   embedded in %%@
their parental molecular cloud frequently show a high fraction of stars with excess emission at %%@
NIR wavelengths.
To investigated the nature of the stellar population in the direction of IRAS15411-5352 field, %%@
we compared  it with the 2MASS photometry of a $ 4.0^\prime \times 4.0^\prime$  nearby field %%@
centered at $\alpha$=15h:44m:40s and $\delta$=-54d:00m:00s, labeled as  "control" in Figure 5, %%@
which probably consists of Galactic field stars. 

We analyzed  the NIR colors of the two selected regions using  the ($J-H$) versus ($H-K$) %%@
comparative diagrams  shown in Figure 6. There, we also represented the position of the main %%@
sequence and the giant branch \citep{koorneef83} as well as the reddening vectors; the location %%@
of 10, 20, 30 and 40 mag of visual extinction are also marked \citep{fitz99}.

In the "control" color-color diagram we can distinguish two stellar groups: the first is formed %%@
by foreground field stars and the other  by cool giants or super-giants, probably located behind %%@
the molecular cloud. In the IRAS15411-5352 color-color diagram, besides the population found in %%@
the control region we can see  other objects showing excess emission at 2.2 $\mu$m, probably due %%@
to the presence of warm circumstellar dust, characteristic of very young  objects %%@
\citep{lada92}. A significant part of this last group is located inside the region labeled %%@
"nebula" in Figure 5.

In order to separate the cluster members  detected only at the $H$ and $K$ bands from background %%@
field stars, we compared the number $N(H-K)$ of objects with colors $H-K$ with that found in the %%@
control population (normalized to the same area). The result is presented in Figure 7, where we %%@
can see that the foreground population is the same in both regions but there is an evident %%@
excess of sources with $H-K$ $\geq$ 0.5 mag in the IRAS15411-5352 region.
Using this result and the $(J-H) \times (H-K)$ diagram constructed from the sources detected %%@
only in the nebula region (shown in Figure 8), we selected as member candidates all sources that %%@
present excess  emission in the NIR, the objects with $H-K \geq 0.5$ that are located  inside %%@
the nebula and do not have $J-H$ colors of late-type stars and also all other sources with  $H-K %%@
\geq 1.5$. 
In Table 1 we present the coordinates and photometry of all selected stars. 

Since the excess emission is generally attributed to the presence of disks or cocoons and  %%@
background giant stars may contaminate the selected sample, we analyzed the recent images
of the nebula made by the Spitzer telescope, which are shown in Figure 9.   
The intensity of the extended emission that we see in the Spitzer images increases with  %%@
wavelength, implying that there is much more dust than what we can infer from the LNA images.
Since Spitzer angular resolution is about 2 times smaller than that of our NIR observations, the %%@
sources inside the nebula region are not resolved, except for IRS1, which appears very bright %%@
and has the same coordinates in both surveys. 
The high extinction at the center of the cluster guarantees that the background contamination  %%@
is small. 

To verify the nature of the LNA selected sources placed at the border of the 
nebula, we constructed their Spectral Enegy Distribution (SED) including the IRAC
data and compared it with the SED of  background red giants. In the late-type sample, the IRAC %%@
flux densities decrease with increasing wavelengths, while the opposite occurs with the cluster %%@
sources, suggesting the presence of  material remaining from the star forming process. We %%@
illustrate this result in Figure 10, where we present the SED of  
IRS8, IRS19 and of late-type star. 
  
IRS7 and IRS16, appear as a single source in the Spitzer images although only IRS16 has excess %%@
emission in the $J-H$ versus $H-K$ diagram.   
IRS8, with  $H-K=4.3$, not detected at LNA $J$ band and located at the west side of the NIR %%@
nebula appears as a bright IRAC source at 3.6 and  4.5$\mu$. The same is valid for
IRS4, which does not show excess  emission at 2.2$\mu$m and therefore can be classified as a %%@
B0-B0.5V ZAMS. 
IRS19 is one of the most interesting of the selected sources. In fact, not only was not detected %%@
at the LNA $J$ band and has $H-K=3.34$, but as we can see from Figure 10, its flux density  %%@
increases continuously from 2.2$\mu$m until 8.0$\mu$m. Finally there is another group of IRAC %%@
sources southeast of the NIR nebula, delimited by the dashed box in Figure 9, from which only %%@
one was detected in the LNA image with magnitude $K=14.95\pm$0.06 and coordinates (J2000) %%@
$\alpha$=15h:45m:04.53s and $\delta$=-54d:02m:39.6s. 

The luminosity of the objects in Table 1 can be derived from the $J$ versus $(J-H)$ %%@
color-magnitude diagram shown in Figure 10 where
the locus of the main-sequence for class V stars at 2.4 kpc \citep{giveon02} is   plotted and %%@
the position of the spectral types earlier than F8V are indicated by labels. 
The intrinsic colors were taken from \citet{koorneef83} while the absolute $J$ magnitudes were %%@
calculated from the absolute visual luminosity for ZAMS taken from \citet{hanson97}. 
The reddening vector for a ZAMS B0 V star, taken from \citet{fitz99}, is shown by the dashed %%@
line with the positions of visual extinction $A_V = 10$, 20 and 30 magnitudes indicated by %%@
filled circles. 
We also indicated the stars with and without excess in the color-color diagram by open and %%@
filled triangles, respectively.
IRS1, the most luminous object in the region,  presents large excess emission in the near %%@
infrared and is the best candidate to the near infrared counterpart of the IRAS source.

\subsection{The IRAS source} \label{bozomath}

In Figure 11 we show the flux calibrated contour diagram made from the $C1$ band image centered %%@
in the IRAS 15411-5352 position with the error ellipse delimited by the broken line, and where %%@
we indicated by filled circles all the cluster member candidates. 
Several sources fall inside the  IRAS error ellipse;  a more accurate coordinate for the IR %%@
source was obtained from the Midcourse Space Experiment - %%@
MSX\footnote{http://www.ipac.caltech.edu/ipac/msx/msx.html} point source catalog. The MSX %%@
surveyed the entire Galactic plane within $\mid b\mid \leq 5^\circ$ in four mid-infrared %%@
spectral bands centered at 8.28, 12.13, 14.65 and 21.34 $\mu$m, with image resolution of 19 %%@
arcsec and a global absolute astrometric accuracy of about 1.9 arcsec \citep{price01}. 
We found one MSX source within the IRAS error ellipse, with coordinates %%@
$\alpha(\rm{J2000)=15^{h}44^{m}59.38^{s}}$, $\delta(\rm{J2000)=-54^{d}02^{m}19.3^{s}}$, which in %%@
fact are coincident with the star that we labeled IRS 1.

In figure 12 we show the spectral energy distribution of the IRS 1 source between 1.25 and 100 %%@
$\mu$m, taken from the IRAS and MSX catalogues and from this work.
We determined a lower limit for the spectral type of IRS 1 using the integrated near to far IR %%@
flux density, assuming that all the stellar luminosity is used to heat the dust that emits at %%@
infrared wavelengths, obtaining $L_{Bol}\geq 7.7\times 10^4 L_\odot$, which corresponds to an %%@
O8.5V ZAMS star \citep{hanson97}.

\subsection{ Br$\gamma$ emission  and the infrared nebula} \label{bozomath}

In order to study the extended emission,  we first scaled the Br$\gamma$ to the  continuum image %%@
($C1$ filter) using the common bright field stars. 
Next we constructed the flux calibrated contour diagram of the difference between the two images %%@
and, by measuring the area between contours, we obtained the net Br$\gamma$ flux using the %%@
relation between magnitude and flux density given in Koorneef (1983). We found  very strong %%@
Br$\gamma$ emission at the position of the IRAS source and more diluted and extended emission %%@
across the continuum nebula. The integrated Br$\gamma$ fluxes are $(9.1\pm 1.2)$ and ($12.6\pm %%@
1.6$) in units of 10$^{-12}$ erg cm$^{-2}$ s$^{-1}$ for the compact and extended sources, %%@
respectively.

In Figure 13 (left side), we show the Br$\gamma$ contour diagram (white lines) together with  %%@
the selected NIR sources (white stars) and  the H$\alpha$ contour diagram taken from Noumaru %%@
$\&$ Ogura (1993) (black lines), overlaying the $C1$ LNA image. In the right side of the Figure %%@
we present the two dimensional Br$\gamma$ intensity distribution, which clearly shows the %%@
compactness of the HII region associated to the IRAS15411-5352 source. In fact, the FWHM of this %%@
distribution is about $1.1^{\prime\prime}$ ($\sim 0.012$ pc at the quoted distance), which %%@
coincides with the stellar PSF, showing that we still did not resolved the region. 

The same procedure used to integrate the  Br$\gamma$ emission was used to calculate the nebular %%@
fluxes at the $J$, $H$ and $K$ bands, resulting in $0.10\pm 0.01$ Jy, $0.29\pm 0.05$ Jy and %%@
$0.38\pm 0.04$ Jy, respectively.
These values can be compared with the expected free-free emission from an optically thin plasma %%@
at wavelength $\lambda$, given by:

\begin{equation}
S_\nu = 5.4\times 10^{-16}T^{-0.5}g_{ff}(\lambda,T)\Omega E e^{-hc/\lambda kT} \; {\rm Jy}
\end{equation}

\noindent
where $E$ is the emission measure (cm$^{-5}$), $\Omega$ the solid angle of the source, $T$ the %%@
temperature in Kelvin and $g_{ff}(\lambda,T)$ the dimensionless  Gaunt factor, which is %%@
approximately 1 for near infrared and for radio can be calculated from \citep{lang78}:

\begin{equation}
g_{ff}(\lambda, T) = \frac{\sqrt{3}}{\pi}\biggl[17.7+\ln
 \biggl(\frac{T^{3/2}\lambda}{c}\biggr)\biggr].
\end{equation}

A lower limit to the product $\Omega E$ can be obtained from the measured Br$\gamma$ fluxes, not %%@
corrected by absorption, using the expression given by \citet{watson97}: 
 
\begin{equation}
S_{\rm Br\gamma}= 0.9h\nu_{\rm Br\gamma}\alpha^{\rm eff}_{\rm Br\gamma}\frac {\Omega}{4\pi}E %%@
\;\; {\rm ergs\; cm^{-2}\; s^{-1}}
\end{equation}

\noindent
where 

\begin{equation}
\alpha^{\rm eff}_{\rm Br\gamma}=6.48\times 10^{-11}T^{-1.06}
\end {equation}

From these $\Omega E$ lower limits and assuming $T=7500$ K, we computed lower limits for  the %%@
free-free emission at the $J$, $H$ and $K$ bands, as well as for the radio emission at 7 mm, %%@
which  are presented in Table 2 together with the observed values. We can see that there are %%@
excess flux densities at the infrared bands, which decreases with the increase in frequency, as %%@
expected in the presence of wavelength dependent absorption. Since the absorption at the %%@
Br$\gamma$ and narrow $K$ bands are equal, the excess in this band can be attributed to thermal %%@
dust emission and/or scattered stellar light.

From the measured sizes (0.12 pc for the extended source and less than 0.02 pc for the compact %%@
one) and the Br$\gamma$ fluxes, we derived emission measures of $5.3\times 10^{24}$ cm$^{-5}$ %%@
and $3.8\times 10^{26}$ cm$^{-5}$, respectively. 
These results are  lower limits because the measured Br$\gamma$ fluxes were not corrected for %%@
extinction and, at least for the compact region,  we only have an upper limit for its size. 
 
Using these values for the emission measure, we calculated the optical depth at radio %%@
wavelengths from the expression:

\begin{equation}
\tau (\lambda, T)=2\times  10^{-23}\;\frac{  g_{ff}(\lambda, T)\lambda^2}{T^{3/2}}E,
\end{equation}

We obtained lower limits for $\tau(7\, \rm mm)$ of the order of 0.02 and 0.003 for the compact %%@
and extended regions, respectively. Notice that the  value obtained for the optical depth of the %%@
extended region can increase by a factor of a few when Br$\gamma$ absorption is included, but it %%@
will still remain much smaller than one. On the other hand, the true optical depth of the %%@
compact region can be much larger than 0.02, if the real size is smaller than our measured %%@
limit. To test this possibility, we derived the emission measure and the corresponding angular %%@
size  under the assumption of $\tau(7\, \rm mm) = 1$. We obtained $E = 1.8 \times 10^{28}$ %%@
cm$^{-5}$ and an angular size of $0.1^{\prime\prime}$, which is about eight times smaller than %%@
our resolution. Assuming large optical depth, the contribution of this region to the total 7 mm %%@
flux density becomes smaller than 10\% of the total flux only if its size is smaller than 2.2 %%@
mas ($2.6 \times 10^{-5}$ pc).

Taking into account these results, we first computed a lower limit to the $K$ band absorption %%@
from the expected and measured $7\, \rm mm$ flux density, assuming optically thin emission for %%@
both sources obtaining $A_K\approx 1.38$ magnitudes, which corresponds to $A_V\sim 14$ mag. 
An upper limit for the absorption was estimated assuming that the compact source is optically %%@
thick at $7\, \rm mm$ and does not contribute to the observed flux, resulting in an absorption %%@
$A_V\approx 20$ mag. The mean extinction of the optical nebula, derived by Noumaru \& Ogura %%@
(1993) from the H$\alpha$/H$\beta$ ratio is 9.8 mag, implying that there is a strong extinction %%@
gradient in the north-south direction.
 
From the observed size of the extended source, the upper limit for the size of the compact %%@
component, the upper and lower limits for the absorption, and assuming spherical symmetry,  we  %%@
estimated  volume  densities of  $(7.4-9.8)\times 10^3$  cm$^{-3}$ and $(1.4-1.9)\times 10^5$ %%@
cm$^{-3}$, representative of compact  (CHII) and ultracompact (UCHII) HII regions %%@
\citep{church02},  respectively.
There is a good agreement between the volume density-size relation ($n_e$ versus $d$) for the %%@
extended source  and what was found for compact and ultra-compact HII regions %%@
\citep{kim01,martin03}, as can be seen in Figure 14. Since we have only an upper limit for the %%@
size of the compact source,  we show in the Figure the track  representing the combinations of %%@
$n_e$ and $d$ that reproduce the observed emission measure. All the points in the track fall %%@
inside the dispersion region defined in \citet{kim01}, which is indicated by the broken lines.   

Finally,  we  estimated  upper and lower limits to the number of Lyman continuum photons %%@
($N_{Ly}$) available in both compact and extended  HII regions, from the corrected Br$\gamma$  %%@
flux densities,  using the expression derived by \citet{ho90}:

\begin{equation}
N_{Ly}= 2.9\times 10^{45} \biggl(\frac{D}{\rm kpc}\biggr)^2 \biggl(\frac {3S_{\rm Br\gamma}}{\rm %%@
10^{-12}\,erg\, cm^{-2}\, s^{-1}}\biggr)\; s^{-1} 
\end{equation}

We obtained $2.2 \times 10^{48} < N_{Ly} < 3.9 \times 10^{48}$ and $1.6 \times 10^{48} < N_{Ly} %%@
< 2.8 \times 10^{48}$ photons s$^{-1}$ for the extended and compact regions, respectively. The %%@
equivalent single star spectral types, obtained from Hanson et al. (1997) are O8.5V-O7.5V, and %%@
O9V-O8.5V, respectively. The spectral type of the star associated to the compact region  agrees %%@
with that  calculated in section 3.2 from the integrated spectral energy distribution of the %%@
IRAS source. However, the spectral type of the star expected to ionize the extended region is %%@
much earlier than what we found for the embedded stars in Figure 3, suggesting that IRS1 is the %%@
single ionizing source of the whole region. Similar results were found by Kim \& Koo (1991) for %%@
other ultracompact cores embedded in more extended HII regions, which they suggested were due to %%@
champagne flow \citep{tenorio79}. From the shape of the observed NIR nebula and the relative %%@
position of the ionizing star (Figure 11), we suggest that this can be also the case of W16-185.

\section{Conclusions} 

We found that the ionized region classified as the planetary nebula W16-185 is in reality %%@
powered by a cluster of young and  massive stars, still embedded in the parental molecular %%@
cloud. From the NIR photometry we identified twenty cluster member candidates; the majority of %%@
them presenting large NIR excess, characteristic of very young objects. 

Using Spitzer infrared images we found that all of selected sources presenting large $H-K$ %%@
colors have bright Spitzer counterparts. From  those images we noted the presence of at least %%@
six highly absorbed mid infrared sources, southeast of the IRS1 source. Only one of them was %%@
detected in the $K$ band of our survey, implying that the extinction must be very high in that %%@
direction.

The IRS1 source is the most luminous object in the region and also presents large infrared %%@
excess. This source is the NIR counterpart of the  MSX source G326.7238+00.6148, which is the %%@
MIR counterpart of the IRAS15411-5352 source. From its near to far spectral energy distribution %%@
and assuming that all the stellar luminosity is used to heat the dust, we determined a lower %%@
limit for its bolometric luminosity of $L_{Bol}\geq 7.7\times 10^4 L_\odot$, which corresponds %%@
to an O8.5V ZAMS star.

We found  very strong Br$\gamma$ emission at the IRAS source position and more diluted and %%@
extended emission across the continuum nebula. The first one is very compact and was not %%@
resolved by our observations, which corresponds to a diameter of less than 0.012 pc at the %%@
distance of 2.4 kpc, size characteristic of ultracompact HII regions. The size of the extended %%@
region is $10^{\prime\prime}$, corresponding to a linear size of 0.11 pc, typical of  compact %%@
HII regions  (Churchwell 2002).

From the observed Br${\gamma}$ flux we computed the emission measure $E$ and the expected %%@
free-free emission at radio wavelengths. From the measured 7 mm flux density, we found lower and %%@
upper limits for the extinction $A_V$ of 14 and 20 mag, respectively. 

From the absorption corrected emission measures, we obtained limits to the volume densities of %%@
$(7.4-9.8)\times 10^3$ and $(1.4-1.9)\times 10^5$ cm$^{-3}$ for the extended and compact %%@
regions, respectively. These density values agree with the assumption of the compact and %%@
ultracompact nature of the sources, as inferred earlier from their sizes. Moreover, the relation %%@
between volume density and size agrees very well with what is found for other compact regions %%@
(eg. Kim \& Koo, 2001).

We also obtained upper and lower limits for the number of Lyman continuum photons necessary to %%@
produce the Br$\gamma$ emission in both the compact and extended region, obtaining $2.2 \times %%@
10^{48} < N_{Ly} < 3.9 \times 10^{48}$ and $1.6 \times 10^{48} < N_{Ly} < 2.8 \times 10^{48}$ %%@
photons s$^{-1}$, respectively. The spectral type of the single ionizing star in the compact %%@
region agrees with that  calculated from the near to far infrared spectral energy distribution. %%@
However, the  stars embedded in the extended region  do not provide enough UV photons to ionize %%@
it, suggesting that IRS1  is probably the main ionizing source of the whole region. In fact, %%@
from the shape of the observed NIR nebula and from the relative position of IRS1 source, we %%@
suggest that it could be produced by the champagne flow as was found for other ultracompact %%@
cores.

%% The \notetoeditor{TEXT} command allows the author to communicate
%% information to the copy editor.  This information will appear as a
%% footnote on the printed copy for the manuscript style file.  Nothing will
%% appear on the printed copy if the preprint or
%% preprint2 style files are used.

%% The eqnarray environment produces multi-line display math. The end of
%% each line is marked with a \\. Lines will be numbered unless the \\
%% is preceded by a \nonumber command.
%% Alignment points are marked by ampersands (&). There should be two
%% ampersands (&) per line.


%% Included in this acknowledgments section are examples of the
%% AASTeX hypertext markup commands. Use \url without the optional [HREF]
%% argument when you want to print the url directly in the text. Otherwise,
%% use either \url or \anchor, with the HREF as the first argument and the
%% text to be printed in the second.

\acknowledgments

This work was partially supported by the Brazilian agencies FAPESP and CNPq.
We acknowledge the staff of Laborat\'orio Nacional de 
Astrof\' \i sica for their efficient support.
This publication makes use of data products from the Two Micron All 
Sky Survey, which is a joint project of the University of 
Massachusets and the Infrared Processing and Analysis Center/California 
Institute of Technology, funded by the
National Aeronautics and Space Administration and the National Science Foundation. 
This research made use of data products from the Midcourse Space 
Experiment. This work is based [in part] on observations made with the Spitzer Space Telescope, %%@
which is operated by the Jet Propulsion Laboratory, California Institute of Technology under a %%@
contract with NASA.

   \begin{figure}
   \centering
   \includegraphics[width=\textwidth]{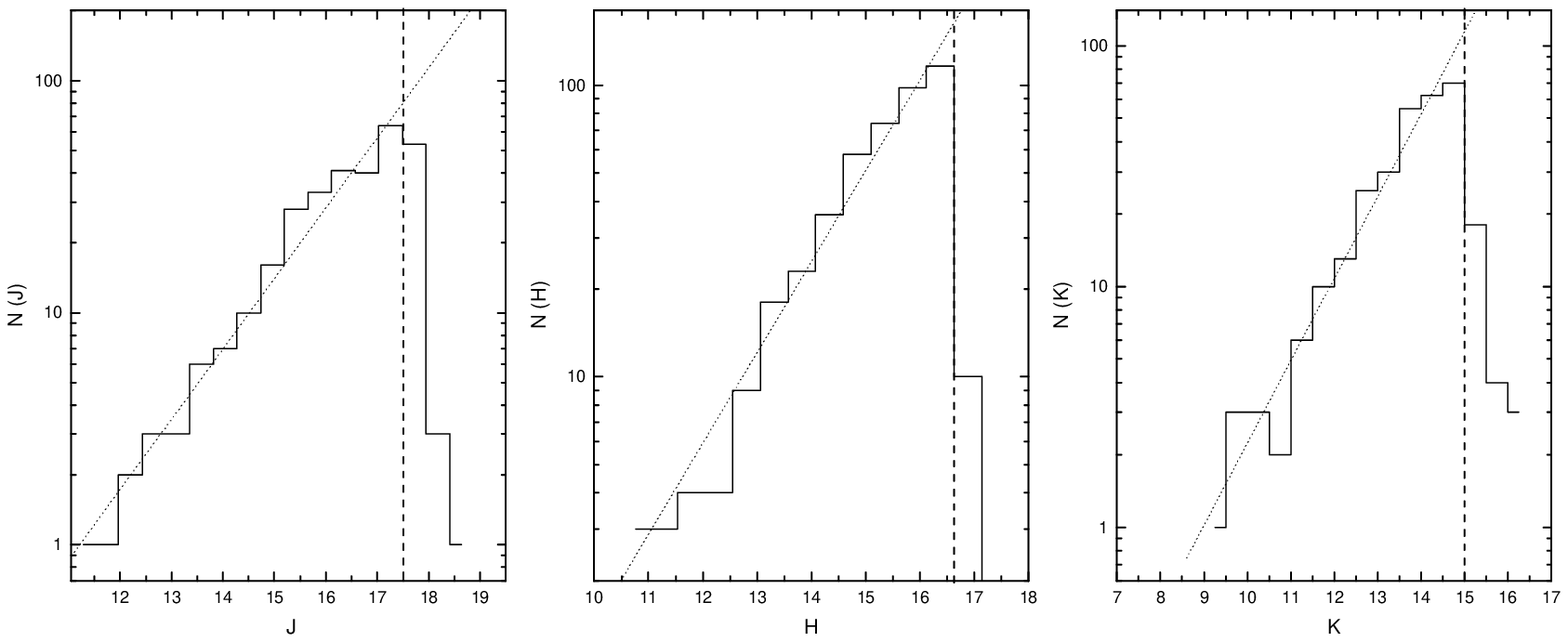}
      \caption{Histograms of $J (left), H (center),$ and $K (right)$ magnitude counts. The %%@
completeness limits values are indicated by the vertical dashed lines.}
         \label{Fig1}
   \end{figure}

\clearpage

   \begin{figure}
   \centering
   \includegraphics[angle=0]{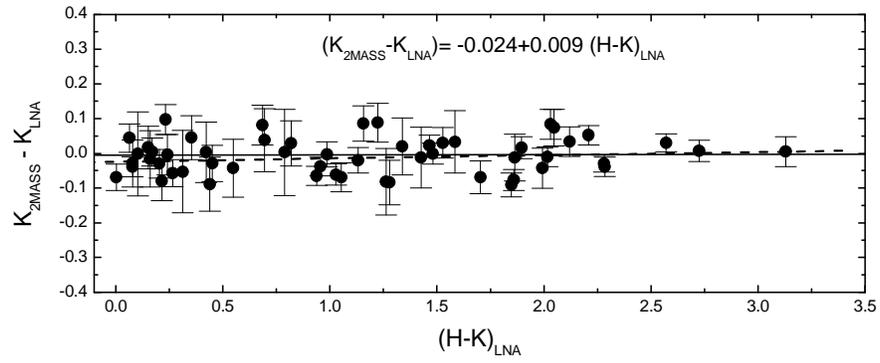}
      \caption{The $K_{2MASS}$- $K_{LNA}$ versus $(H-K)_{LNA}$ diagram. The dashed line %%@
represents the linear fit to the data, which is shown on the top. We can see that the color term %%@
correction is very small.}
         \label{Fig2}
   \end{figure}

   \begin{figure}
   \centering
   \includegraphics[angle=0]{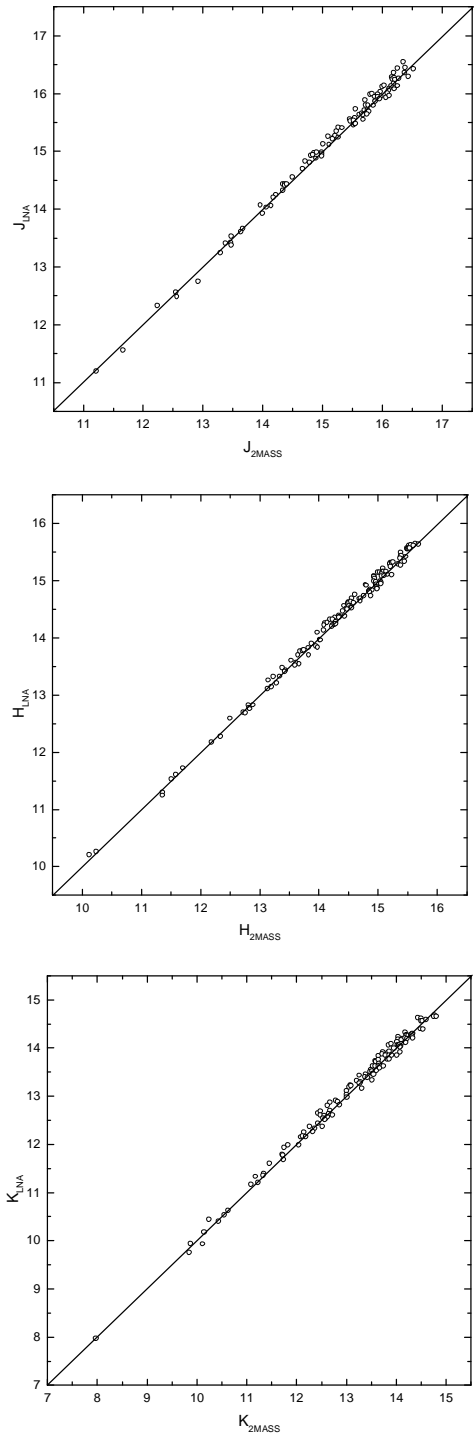}
      \caption{The LNA versus 2MASS comparative diagrams. We compared 101 ($J$), 131 ($H$) and %%@
126 ($K$) stars in both surveys in an area of about 13 square arcmin around IRAS15411-5352. The %%@
straight lines represent the expected relation if the two photometric system were equal.}
         \label{Fig3}
   \end{figure}

\begin{figure*}
   \centering
   \includegraphics[width=\textwidth]{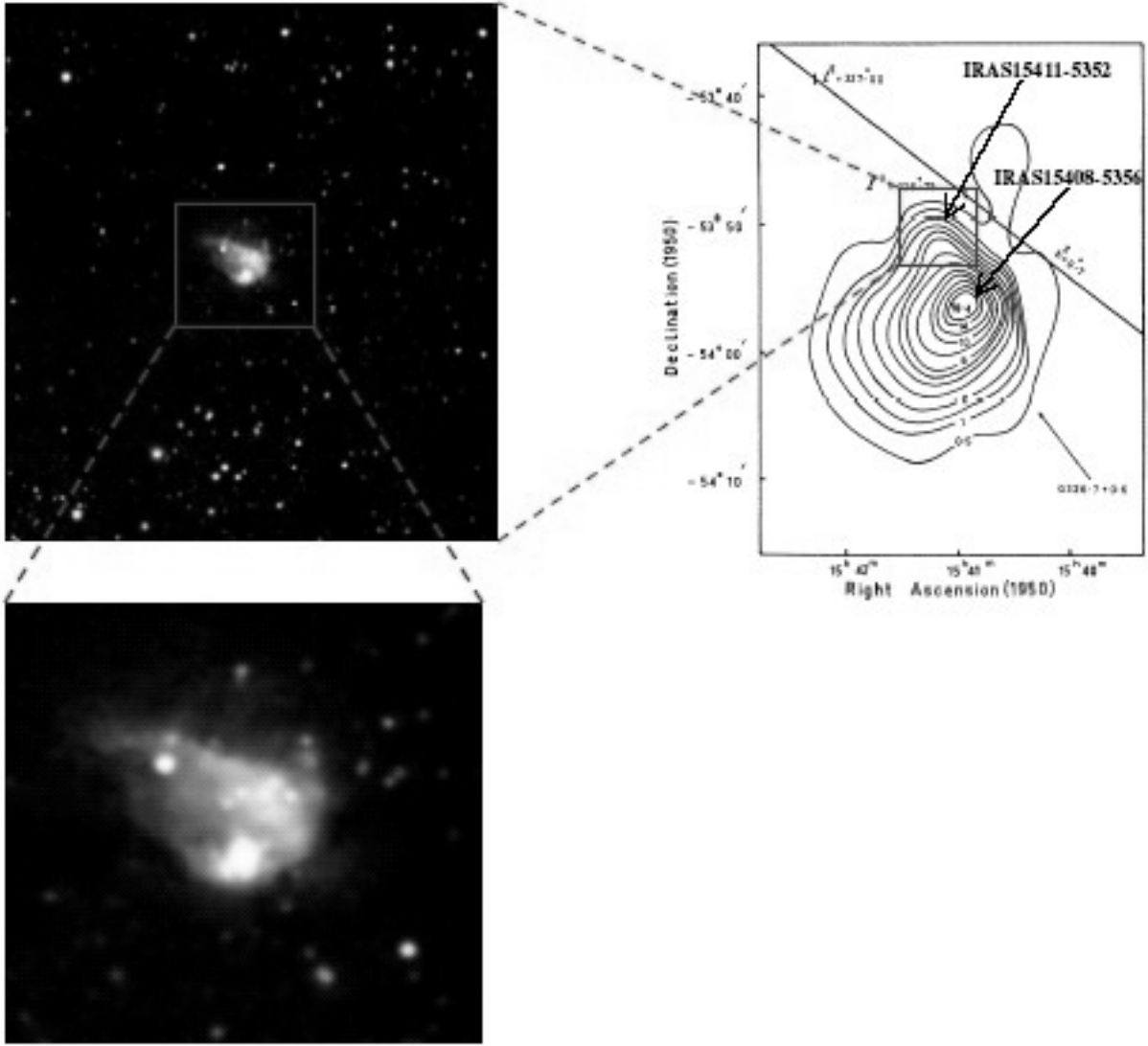}
      \caption{Combined false-image (3.9 $\times$ 3.5 square arcmin) made from the $J$ (blue), %%@
$H$ (green) and 
nb$K$ (red) LNA images. North is to the top east to the left. At the botton-left side we present %%@
a detailed view (about 1.2 $\times$ 1.0 square arcmin) of the near infrared nebula. The 5 GHz %%@
contour map (Goss \& Shaver, 1970) of the region, is shown in the right side of the Figure, %%@
where  the location of IRAS 15408-5356 and IRAS 15411-5352 sources are indicated.The IRAS %%@
15411-5352 source is associated with a small deformation in the 5 GHz contour map.}
         \label{Fig4}
   \end{figure*}

   \begin{figure*}
   \centering
   \includegraphics[width=\textwidth]{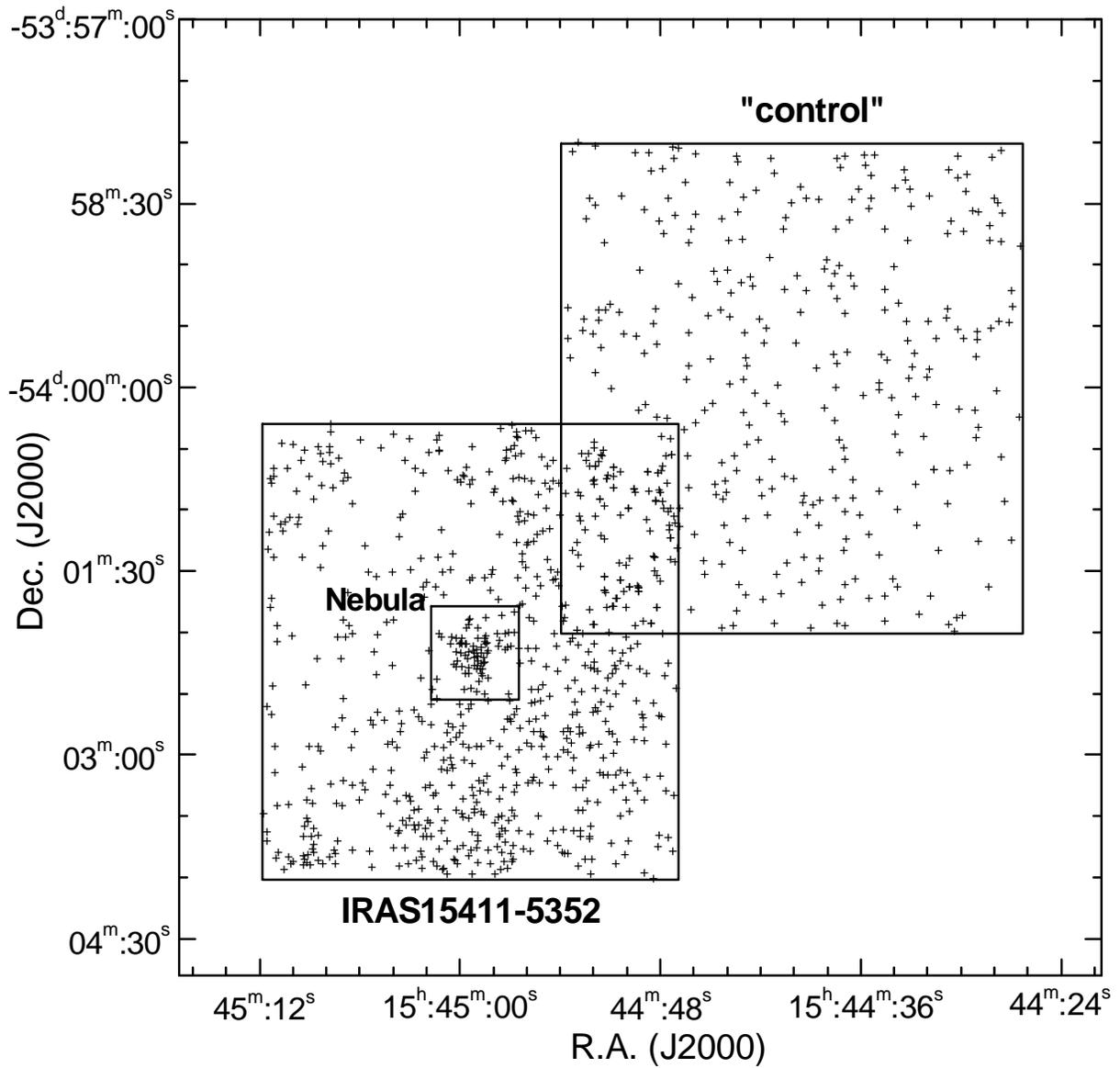}
      \caption{Diagram of spatial distribution of sources detected in the LNA $K$ band image %%@
(IRAS15411-5352) and for the "control" region taken from the 2MASS survey. The Nebula region is %%@
indicated by the small box inside the IRAS region.}
         \label{Fig5}
   \end{figure*}

   \begin{figure}
   \centering
   \includegraphics[angle=0]{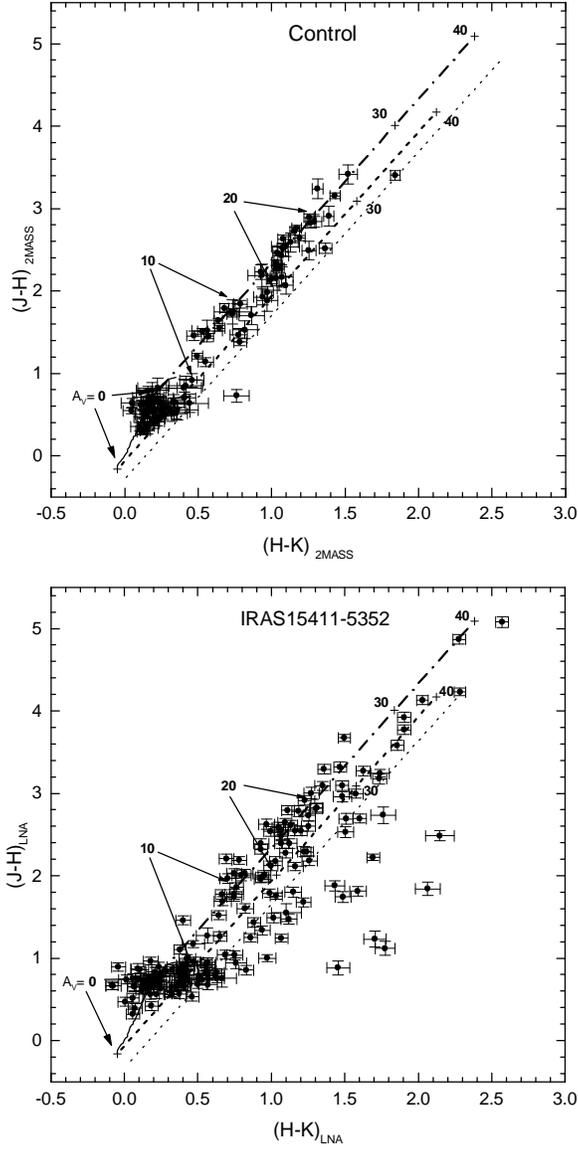}
      \caption{The $(J-H)$ versus $(H-K)$ comparative diagrams. The locus of the main sequence
and giant branch taken from Koornneef (1983), are shown by the continuous lines,
while the two parallel lines (dashed and dotted) follow the reddening vectors given by %%@
Fitzpatrick (1999).
The crosses represent the location of $A_V$ = 0, 10, 20, 30 and 40 magnitudes of visual %%@
extinction, which are indicated by bold numbers. The dotted line separate the NIR sources that %%@
present excess emission from that do not.}
         \label{Fig6}
   \end{figure}

      \begin{figure}
   \centering
   \includegraphics[angle=0]{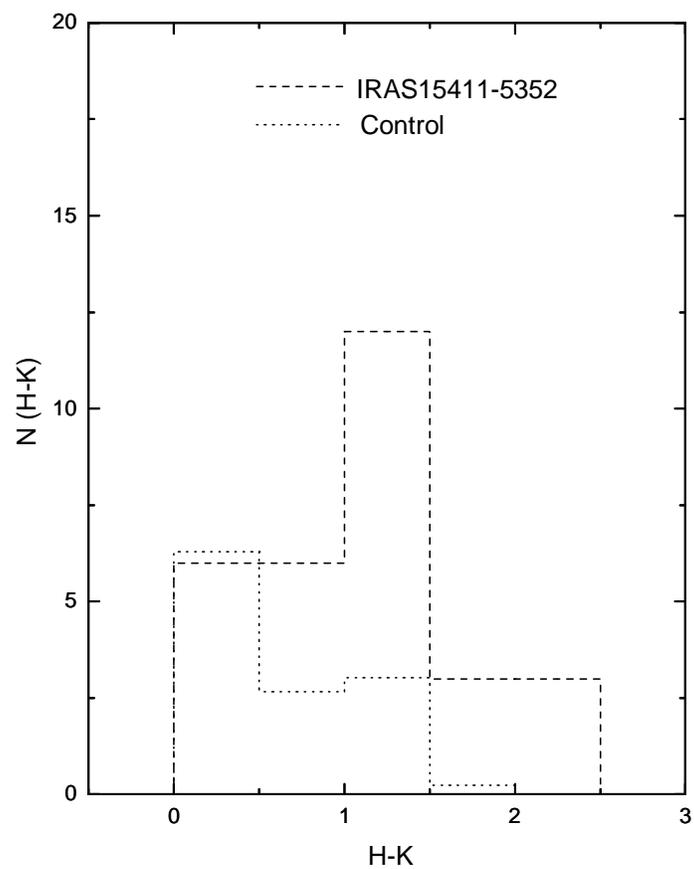}
      \caption{The $N(H-K)$  comparative diagram for the control (dotted line) and Nebula %%@
(dashed line) regions. The counts from the Control and Nebula region were normalized to the same %%@
area.}
         \label{Fig7}
   \end{figure}

       \begin{figure}
   \centering
   \includegraphics[angle=0]{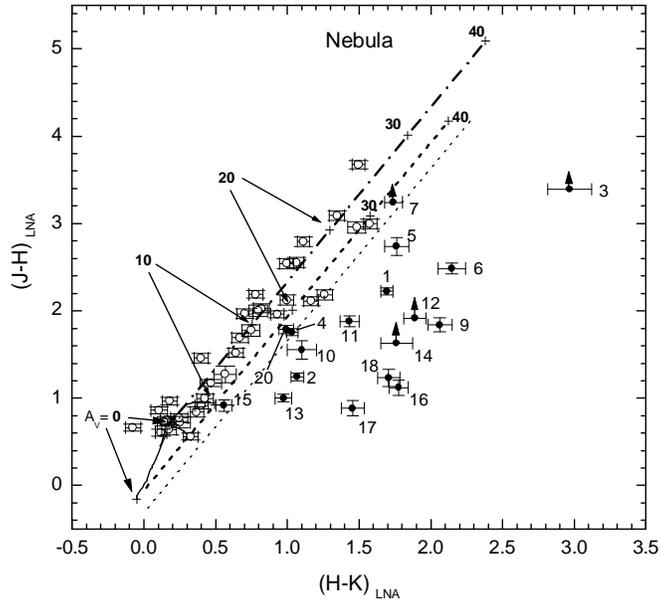}
      \caption{The $(J-H)$ versus $(H-K)$ diagram for the nebula region. The locus of the main %%@
sequence and giant branch taken from Koornneef (1983), are shown by the continuous lines,
while the two parallel lines (dashed and dotted) follow the reddening vectors given by %%@
Fitzpatrick (1999).
The crosses represent the location of $A_V$ = 0, 10, 20, 30 and 40 magnitudes of visual %%@
extinction, which are indicated by bold numbers. The cluster member candidates were selected %%@
from the criteria described in the text. The dotted line separate the NIR sources that present %%@
excess emission from that do not. Some sources detected only in the $H$ and $K$ bands were %%@
plotted using the $J$ band completeness magnitude limit (represented by arrows).}
         \label{Fig8}
   \end{figure}   
   
  \begin{figure}
   \centering
\plotone{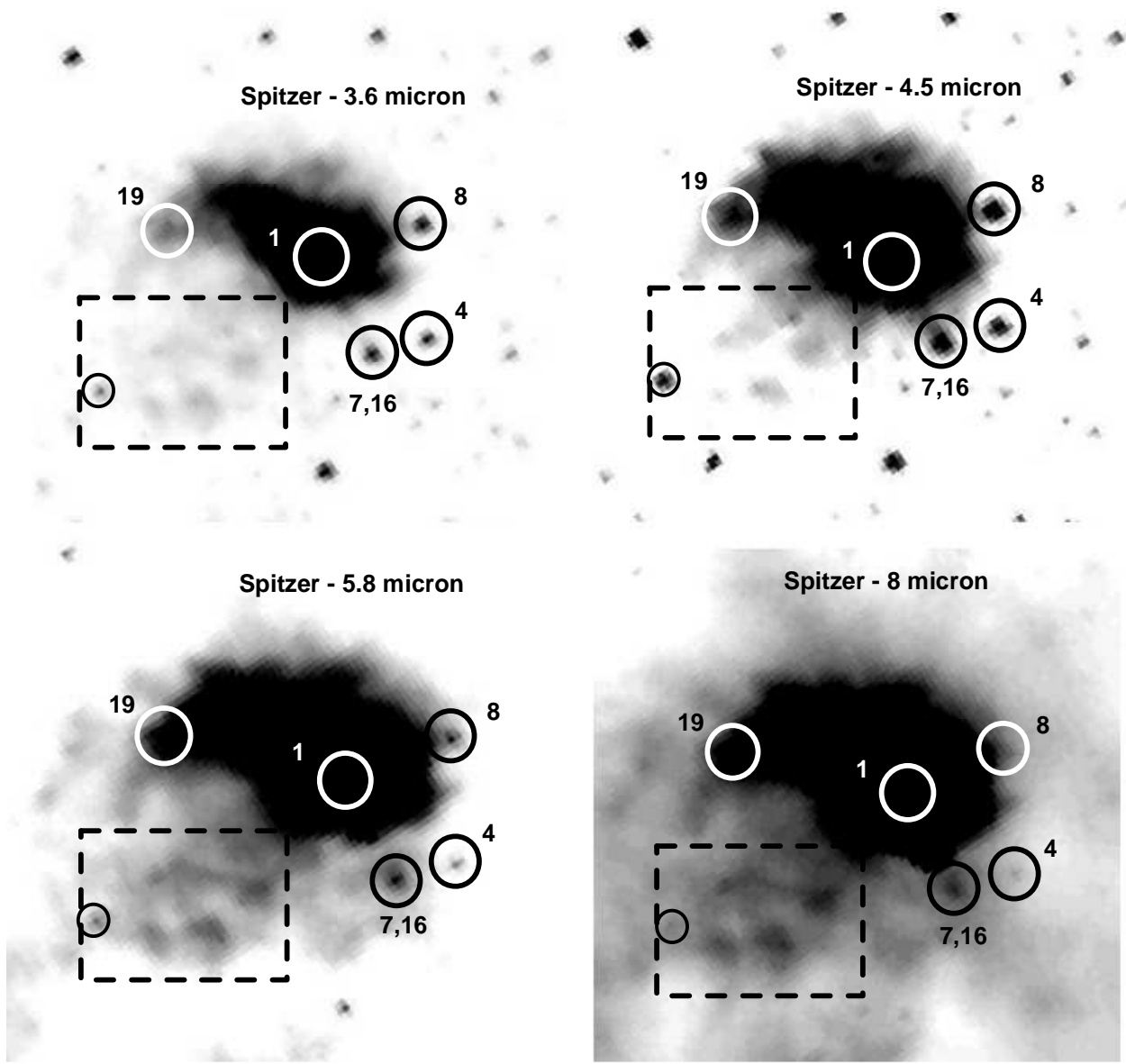}   
      \caption{ The 1.4 $\times$ 1.0 arcmin images taken from the Spitzer data archive (North is %%@
to the top, east to the left) of the region around IRAS15411-5352. The Infrared Array Camera %%@
(IRAC) has four infrared bands centered at 3.6, 4.5, 5.8 and 8.0$\mu$m with image plate scale of %%@
1$^{\prime\prime}$.2. Some prominent LNA sources are indicated by labels. The small dashed box %%@
in the southeast of the images delineate a group of highly absorbed mid-infrared sources. The %%@
small black circle indicate the only one in this group detected in our $K$ band image.}
         \label{Fig9}
   \end{figure} 
   
     \begin{figure}
   \centering
\plotone{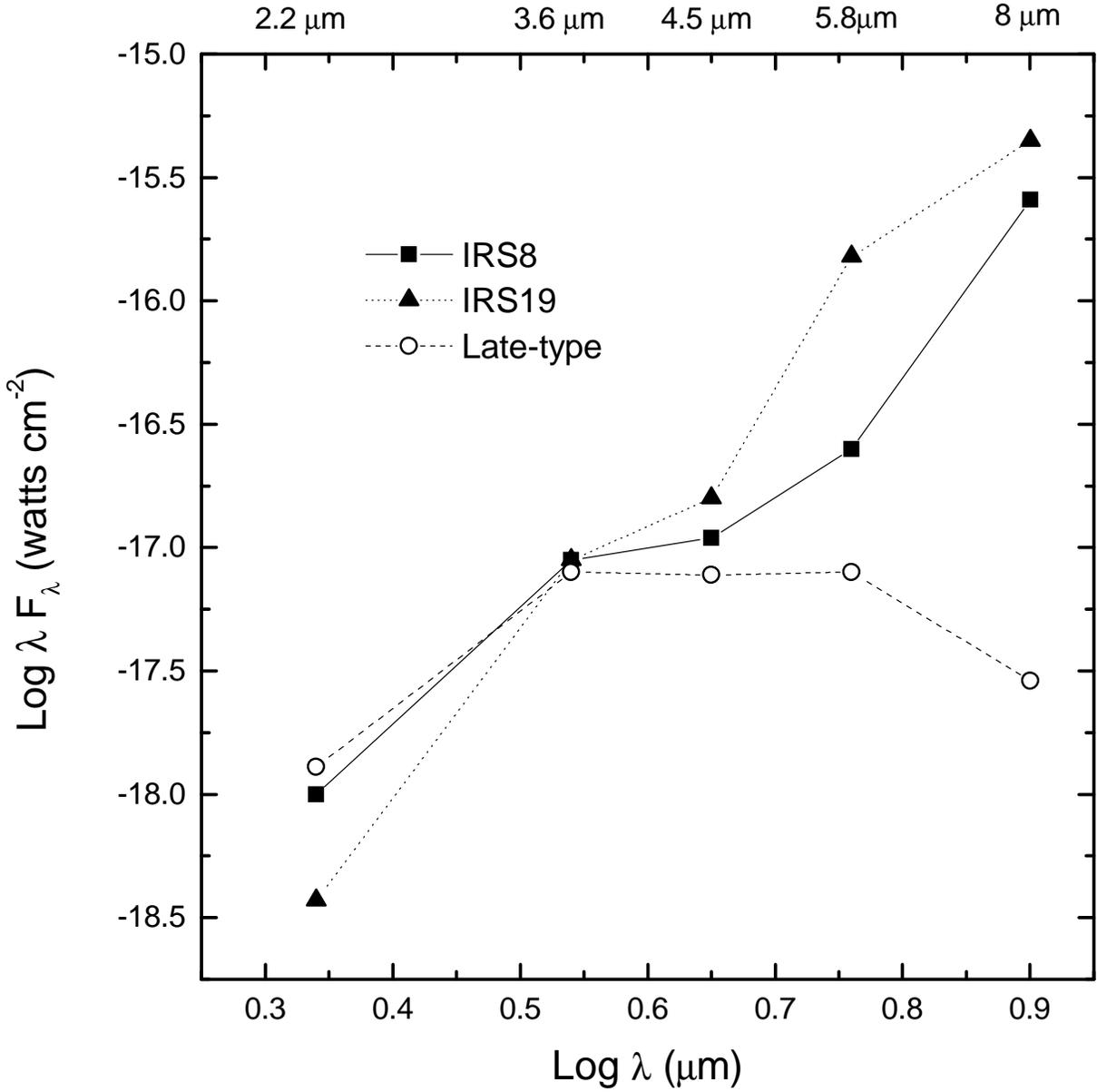}   
      \caption{ The Spectral Energy Distribution (between 2.2 and 8.0$\mu$m) of the IRS8 (filled %%@
squares) and IRS19 sources (filled triangles), together with the SED of a late-type star (open %%@
circles). The 2.2$\mu$m flux densities were taken from the LNA survey; the 3.6-8.0$\mu$m data %%@
were taken from IRAC.}
         \label{Fig10}
   \end{figure}

 \begin{figure}
   \centering
   \includegraphics[angle=0]{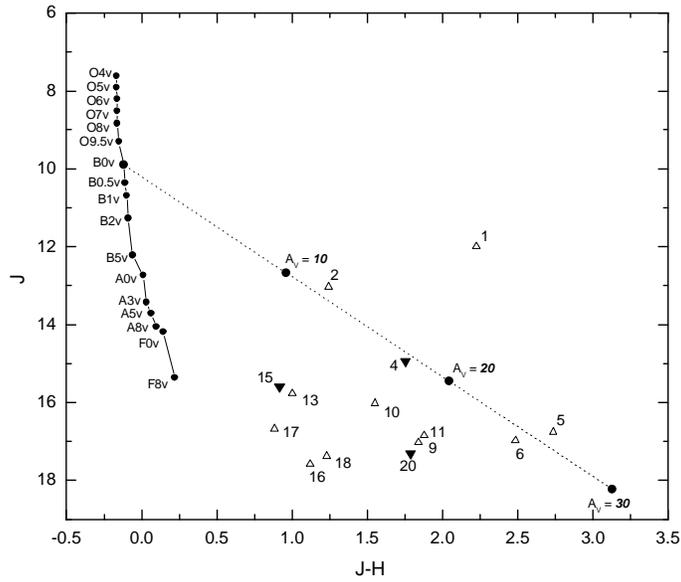}
      \caption{The $J$ versus $(J-H)$ color-magnitude diagram of the sources in Table 1.
The locus of the main sequence at 2.4 kpc (Giveon et al. 2002) is shown by the continuous line. %%@
The intrinsic colors were taken from Koornneef (1983) while the absolute $\it{J}$ magnitudes %%@
were calculated from the absolute visual luminosity of ZAMS taken from
Hanson et al. (1997). The reddening vector for a B0V star (dotted line) was taken
from Fitzpatrick (1999). The bold numbers indicate the location  of $A_V$ = 10, 20 and 30 %%@
magnitudes of visual extinction. Open up triangles represent the  sources that show excess %%@
emission in the NIR and filled down triangles those that do not.}
         \label{Fig11}
   \end{figure}

  \begin{figure}
   \centering
\plotone{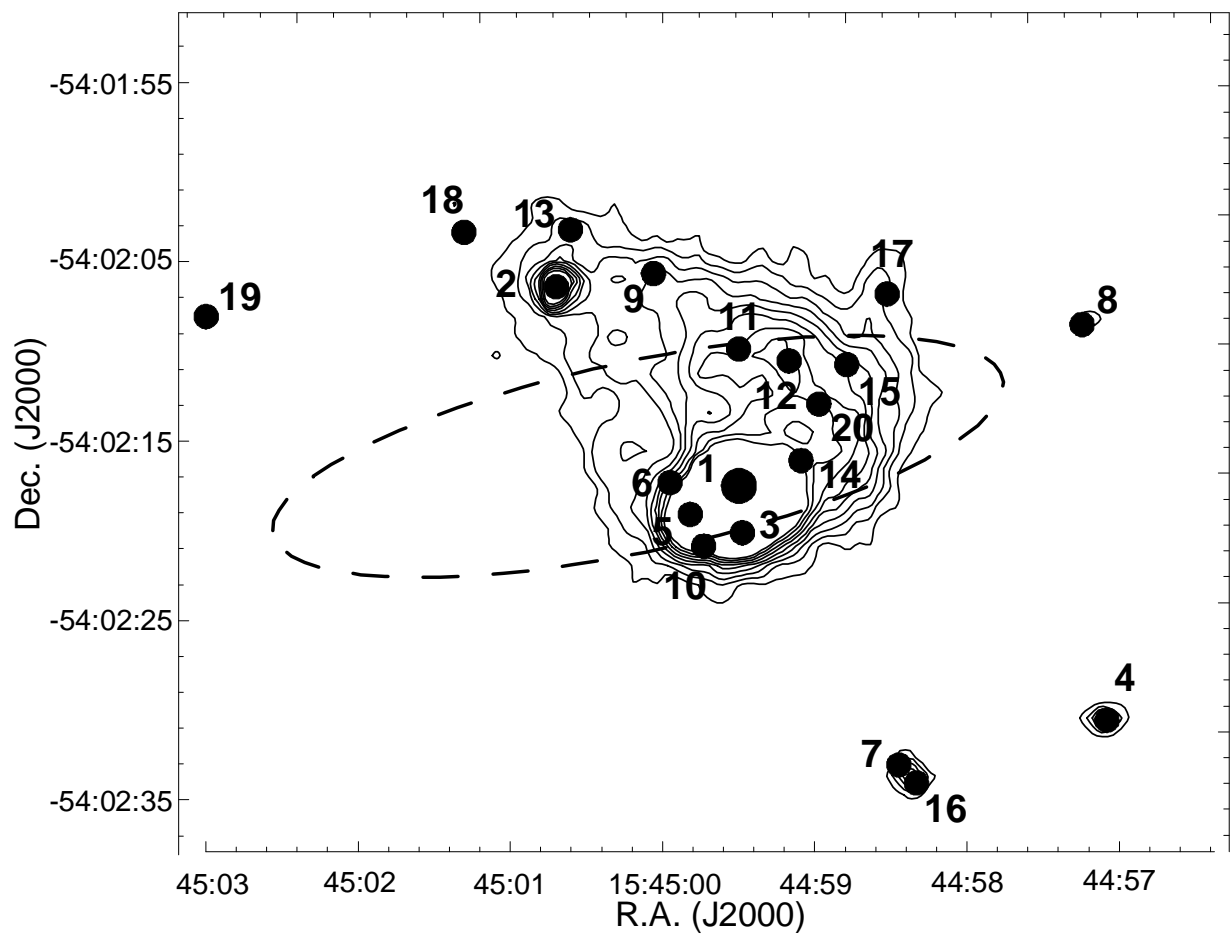}   
      \caption{$K$ band contour map of the infrared nebula.
The contours start at $3.5\times 10^{-5}$ Jy arcsec$^{-2}$ and are spaced by $2.1\times %%@
10^{-5}$Jy arcsec$^{-2}$. 
The positions (points) and labels (numbers) of the selected infrared sources referred in Table 1 %%@
are indicated as well as the position of the IRAS coordinate error ellipse (dashed line).}
         \label{Fig12}
   \end{figure}

     \begin{figure}
   \centering
\plotone{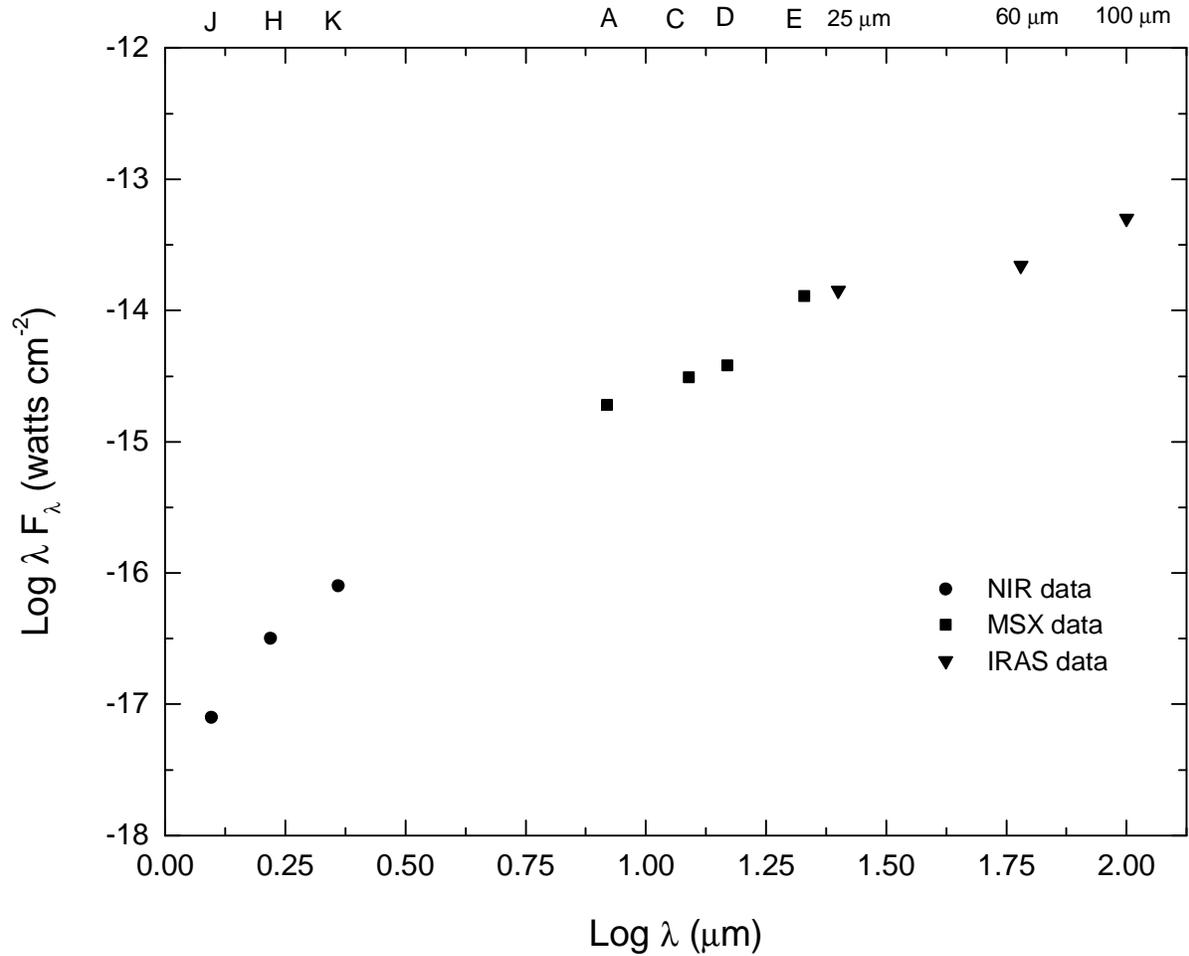}
      \caption{The spectral energy distribution of the IRS1 infrared source. The flux for the %%@
three near infrared bands (filled circles) were taken from our work. The mid infrared data %%@
(filled squares) were taken from $MSX$ catalogue
(bands A=8.28$\mu$m, C=12.13$\mu$m, D=14.65$\mu$m and E=21.34$\mu$m) while the far infrared data %%@
(filled down triangles) were taken from the IRAS point source catalogue.}
         \label{Fig13}
   \end{figure}

 \begin{figure*}
   \centering
   \plotone{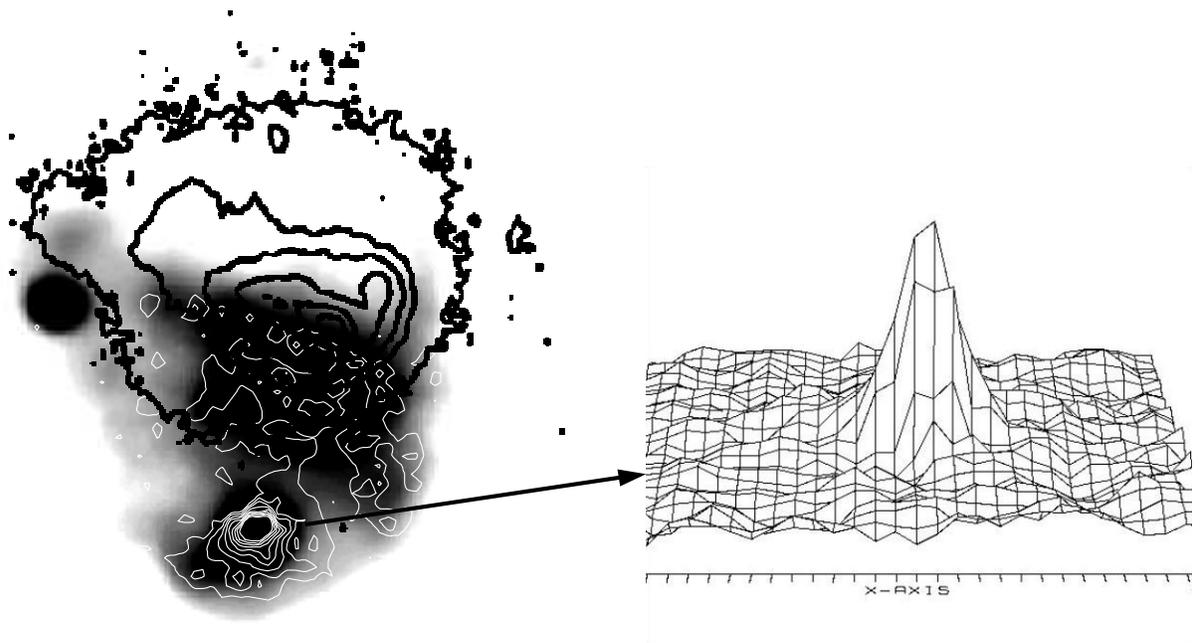}
      \caption{The LNA $K$ band image (about 0.45 $\times$ 0.5 square arcmin) of the studied %%@
region, superimposed by the Br$\gamma$ (white lines) and H$\alpha$ (black lines) (given by %%@
Noumaru $\&$ Ogura, 1993), contour diagrams. The Br$\gamma$ contours are flux calibrated, with %%@
the values starting at $1\times 10^{-6}$ Jy arcsec$^{-2}$ and spaced by $7.5\times 10^{-7}$Jy %%@
arcsec$^{-2}$. 
The two dimensional intensity distribution of the Br$\gamma$ point source, is presented at the %%@
right side of the Figure.}
         \label{Fig14}
   \end{figure*}

     \begin{figure}
   \centering
\plotone{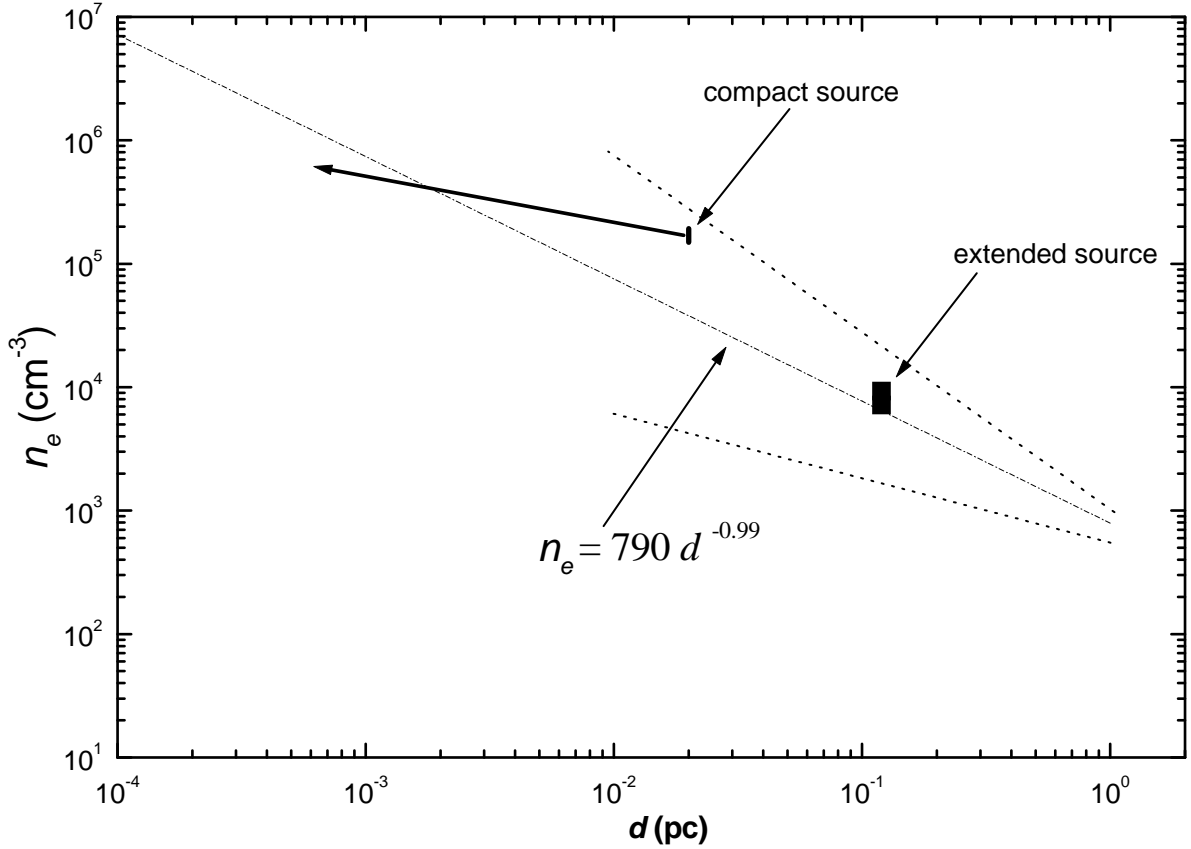}   

      \caption{The observed relation between volume electron density $n_e$ and size $d$ for the %%@
compact (filled rectangle) and ultracompact (continuous line) HII regions. The continuous black %%@
bold line represents the positions of all points that give the observed emission measure for the %%@
unresolved $Br\gamma$ source. The traced line represents the mean relation obtained by Kim \& %%@
Koo (2001) for other compact HII regions, and the dotted line, the dispersion of the %%@
observations from which the mean relation was obtained.}
         \label{Fig15}
   \end{figure}

\clearpage

\begin{deluxetable}{crrrrrr}
\tabletypesize{\scriptsize}
\tablecaption{List of the selected near-infrared sources \label{tbl-1}}
\tablewidth{0pt}
\tablehead{
\colhead{IRS}&\colhead{$\alpha$(J2000)}&\colhead{$\delta$(J2000)}&
\colhead{$J$}&
\colhead{$H$}&\colhead{$K$}&
}
\startdata

\hline

1	&	15	:	44	:	59.46	&	-54	:	02	:	17.4	&	11.99	$\pm$	0.02	\phn&	%%@
9.77	$\pm$	0.03	\phn&	8.07	$\pm$	0.03	\phn&\\
2	&	15	:	45	:	00.60	&	-54	:	02	:	05.6	&	13.02	$\pm$	0.03	\phn&	%%@
11.78	$\pm$	0.03	\phn&	10.71	$\pm$	0.03	\phn&\\
3	&	15	:	44	:	59.42	&	-54	:	02	:	18.6	&			\phn&	14.11	$\pm$	%%@
0.12	\phn&	11.14	$\pm$	0.10	\phn&\\
4	&	15	:	44	:	57.08	&	-54	:	02	:	30.4	&	14.94	$\pm$	0.02	\phn&	%%@
13.19	$\pm$	0.03	\phn&	12.15	$\pm$	0.03	\phn&\\
5	&	15	:	44	:	59.76	&	-54	:	02	:	18.9	&	16.76	$\pm$	0.07	\phn&	%%@
14.02	$\pm$	0.08	\phn&	12.26	$\pm$	0.07	\phn&\\
6	&	15	:	44	:	59.90	&	-54	:	02	:	17.2	&	16.97	$\pm$	0.05	\phn&	%%@
14.49	$\pm$	0.05	\phn&	12.34	$\pm$	0.09	\phn&\\
7	&	15	:	44	:	58.35	&	-55	:	02	:	33.3	&				\phn&	14.26	%%@
$\pm$	0.05	\phn&	12.52	$\pm$	0.04	\phn&\\
8	&	15	:	44	:	57.14	&	-54	:	02	:	07.5	&				\phn&	17.38	%%@
$\pm$	0.08	\phn&	13.00	$\pm$	0.06	\phn&\\
9	&	15	:	44	:	59.95	&	-54	:	02	:	04.9	&	17.03	$\pm$	0.07	\phn&	%%@
15.19	$\pm$	0.04	\phn&	13.12	$\pm$	0.07	\phn&\\
10	&	15	:	44	:	59.69	&	-54	:	02	:	19.9	&	16.02	$\pm$	0.07	\phn&	%%@
14.46	$\pm$	0.08	\phn&	13.36	$\pm$	0.06	\phn&\\
11	&	15	:	44	:	59.45	&	-54	:	02	:	09.6	&	16.84	$\pm$	0.05	\phn&	%%@
14.96	$\pm$	0.04	\phn&	13.53	$\pm$	0.05	\phn&\\
12	&	15	:	44	:	59.14	&	-54	:	02	:	10.0	&			\phn&	15.59	$\pm$	%%@
0.05	\phn&	13.70	$\pm$	0.06	\phn&\\
13	&	15	:	45	:	00.48	&	-54	:	02	:	02.4	&	15.77	$\pm$	0.05	\phn&	%%@
14.77	$\pm$	0.06	\phn&	13.80	$\pm$	0.06	\phn&\\
14	&	15	:	44	:	59.10	&	-54	:	02	:	16.1	&			\phn&	15.87	$\pm$	%%@
0.09	\phn&	14.11	$\pm$	0.06	\phn&\\
15	&	15	:	44	:	58.76	&	-54	:	02	:	10.1	&	15.60	$\pm$	0.06	\phn&	%%@
14.68	$\pm$	0.04	\phn&	14.12	$\pm$	0.05	\phn&\\
16	&	15	:	44	:	58.28	&	-54	:	02	:	33.8	&	17.57	$\pm$	0.09	\phn&	%%@
16.46	$\pm$	0.13	\phn&	14.27	$\pm$	0.06	\phn&\\
17	&	15	:	44	:	58.47	&	-54	:	02	:	05.3	&	16.67	$\pm$	0.06	\phn&	%%@
15.79	$\pm$	0.06	\phn&	14.33	$\pm$	0.06	\phn&\\
18	&	15	:	45	:	01.12	&	-54	:	02	:	02.3	&	17.37	$\pm$	0.09	\phn&	%%@
16.14	$\pm$	0.06	\phn&	14.44	$\pm$	0.05	\phn&\\
19	&	15	:	45	:	02.90	&	-54	:	02	:	07.5	&				\phn&	17.79	%%@
$\pm$	0.09	\phn&	14.45	$\pm$	0.08	\phn&\\
20	&	15	:	44	:	59.01	&	-54	:	02	:	12.2	&	16.75	$\pm$	0.09	\phn&	%%@
15.99	$\pm$	0.07	\phn&	15.33	$\pm$	0.12	\phn&\\

\hline

\enddata

%% Text for table notes should follow after the \enddata but before
%% the \end{deluxetable}. Make sure there is at least one \tablenotemark
%% in the table for each \tablenotetext.

%tablenotetext{* }{Stars that show "excess" of emission at 2.2 $\mu$m }

\end{deluxetable}

  \begin{table}
 \begin{minipage}[t]{\columnwidth}
 \caption{List of derived parameters \label{tbl-3}}
%\tabletypesize{\scriptsize}
 \begin{tabular}{lllllllll@{}llllllllllll@{}llllllllllll@{}llllll@{}}
 \hline
Parameter& extended & compact & total &
\\
\hline

Angular diameter ($"$)& 10	& $< 1.9  $& & \\
Radius (pc)& 0.06& $< 0.01$& & \\
S$_{Br\gamma}$(obs)\footnote{the Br$\gamma$ fluxes are in units of 10$^{-12}$erg s$^{-1}$}& %%@
$12.6\pm 1.6$	& $9.1\pm 1.2$ & $21.7 \pm 2.0$& \\
%S$_{Br\gamma}$(pred)& 	&	& & \\
S$_{43 GHz}$(obs)\footnote{the remaining fluxes are in Jy} & 	&	& $5.2\pm 0.9$ & \\
S$_{43 GHz}$(pred)& 0.85	& 0.61	& 1.46 & \\
S$_{J}$(obs)& 	&	& 0.17$\pm 0.02$& \\
S$_{J}$(free-free)& 	&	& 0.10& \\
S$_{H}$(obs)& 	&	& 0.29$\pm 0.05$& \\
S$_{H}$(free-free)& 	&	& 0.14& \\
S$_{K}$(obs)& 	&	& 0.38$\pm 0.06$& \\
S$_{K}$(free-free)& 	&	& 0.18& \\
$\tau (43)$&  $> 0.003$& $>0.02$& & \\
$E_{Br_\gamma}(\rm cm^{-5}$)& $> 5.3\times 10^{24}$& $> 3.8\times 10^{26}$& & \\
$n_e (\rm cm^{-3})$& 7.4 -- 9.8 $\times 10^{3}$&1.4 -- 1.9 $\times 10^{5}$& & \\
$A_V$&$$&$$&$14-20$& \\
$N_{Ly}\footnote{in units of $10^{48}$s$^{-1}$}$&$2.2-3.9$&$1.6-2.8$&$3.8-6.7$& \\
\hline
\end{tabular}
\end{minipage}
\end{table}

\end{document}